# PdTe a 4.5K Type II BCS Superconductor

Brajesh Tiwari[1], Reena Goyal[1], Rajveer Jha[1], Ambesh Dixit[2] and V. P. S Awana[1,*]

[1]CSIR-National Physical Laboratory, Dr. K. S. Krishnan Marg, New Delhi-110012, India
[2]Center for Energy, Indian Institute of Technology Jodhpur, Rajasthan 342011, India

**Abstract** – We report on the structure and physical properties of polycrystalline PdTe superconductor, which is synthesized by solid state reaction route, via quartz vacuum encapsulation technique at 750°C. The as synthesized compound is crystallized in hexagonal crystal structure with in P63/mmc space group. Both transport and magnetic measurements showed that PdTe is bulk superconductor below 4.5K. Isothermal magnetization (MH) and Magneto-transport {R(T)H} measurements provided the values of lower ($H_{c1}$) and upper ($H_{c2}$) critical field to be 250Oe and 1200Oe respectively at 2K, establishing that the compound is clearly a type-II superconductor. The Coherence length ($\xi_0$) and Ginzburg–Landau parameter ($\kappa$) are estimated from the experimentally determined upper and lower critical fields, which are 449Å and 1.48 respectively. Thermodynamic heat capacity measurements under different magnetic fields, i.e. $C_p(T)H$, showed clear transition at 4.5K ($T_c$), which shifts gradually to lower temperatures with application of field. The values of Debye temperature ($\Theta_D$) and electronic specific heat coefficient ($\gamma$) being obtained from $C_p(T)$ data are found to be 203K and 6.01mJ/mole-$K^2$. The observed specific heat jump ($\Delta C/\gamma T_c$) is 1.33, thus suggesting possible weak coupling case for PdTe superconductor.

PACS 74.70.-b Superconducting materials other than cuprates

PACS 74.25.-q Properties of superconductors

PACS 74.25.Bt Thermodynamic properties

*Corresponding Author
Dr. V. P. S. Awana, Principal Scientist
E-mail: awana@mail.npindia.org
Ph. +91-11-45609357, Fax-+91-11-45609310
Homepage awanavps.webs.com

**Introduction:**

Search for new superconductors has been active research topic of experimental and theoretical condensed matter physics for over a century due to their extraordinary physical properties and potential for applications. The recent discovery of superconductivity in Fe-based compounds, especially doped and undoped iron-chalcogenide rejuvenated the interest, as it contests the conventional thinking against magnetism in superconductors [1-4]. More recently superconductivity in 4d and 5d transition metal based compound like $(Ta/Nb)_2Pd_x(S/Te)_5$ has been getting much attention due to their high critical field, being outside the Pauli paramagnetic limit [5-7]. The strong hybridization of d and p orbital of Pd and Te respectively, dominates the strong covalence, which along with lack of orbital degeneracy, results in much reduced strong correlations. Ekuma et al. demonstrated strong three dimensional character of Fermi- surface in PdTe, which rules out the possibility of high-Tc superconductivity in contrast to FeSe [8]. As far as the superconductivity of PdTe is concerned, the search and understanding of superconductivity in non-superconducting elements containing compounds had been of much interest for years to many researchers [9, 10]. Way back in 1953, Mathias listed PdTe with $T_c$=2.3K for the first time [9]. The phase diagram of PdTe-$PdTe_2$ was later reported with their magnetic and electrical properties by Kjekshus et al., [11]. It is interesting to note, that it took around sixty years before superconductivity of PdTe could once again be tracked in detail by Karki et al., in 2012 [12] and extended their work by doping Fe on Pd site and complete the phase diagram [13]. Interestingly, Karki et al., found a $T_c$ of 4.5K for single crystalline PdTe samples [12], which is nearly two times to that as reported in 1953 by Mathias [9]. Keeping in view that PdTe superconductivity is only scant in literature [9-14] and that too with very different $T_c$ values of 2.3K [9] and 4.5K [12]. Karki et al obtained small shiny crystals (15µm) from high temperature $1080^0$C melt of constituent PdTe and further slow cooling of the same at a rate of $5^0$C/hour. The small crystallites were possibly taken from the bulk of the melt ingots. In present work we report on the synthesis and superconductivity of polycrystalline PdTe compound. The polycrystalline PdTe is obtained by heating the mixed constituent elements at $750^oC$ with a rate of $2^0$C/min for 24h and subsequently cooled to room temperature. Such obtained polycrystalline sample exhibited superconductivity at above 4.5K. Various superconducting critical parameters, obtained experimentally by transport, magnetic, and thermodynamic measurements are reported here.

**Methods:**

Polycrystalline bulk PdTe compound was synthesized via solid state route. The constituent elements Pd (99.9%-3N) and Te (99.99%-4N) from Sigma Aldrich are mixed in a stoichiometry ratio of 1.1:1 in argon controlled glove box and then pelletized by applying uniaxial stress of 100kg/cm$^2$. The pellet sealed in an evacuated ($< 10^{-3}$ Torr) quartz tube was kept in a furnace immediately for heating at 750$^o$C with a rate of 2$^0$C/min for 24h. Thus obtained sample was dense, shiny black and in one piece. For different physical property measurements, the sample was broken into desired pieces. A part of the as synthesized sample is shown along side in Figure 1. The present sample is polycrystalline, obtained by normal heating at 750$^o$C and is different than the tiny single crystals obtained by Karki et al. [12], from the high temperature melt (1080$^o$C) of constituent elements. The structural characterization was done with Rigaku x-ray diffractometer using CuKα radiation of 1.5418Å. Electrical and magnetic measurements were performed on Quantum Design (QD) Physical Property Measurement System (PPMS) - down to 2K.

**Results and discussion:**

The observed powder XRD pattern is Rietveld refined for hexagonal structure with space group P6$_3$/mmc (#129) using Full-Prof and is shown in Figure 1 for as synthesized PdTe. The global fitness of pattern is $\chi^2 = 2.84$, which is basically the mean square of the difference between the experimentally observed and Rietveld fitted XRD patterns. The obtained lattice parameters are a=b=4.15331(15)Å, c=5.6732(4)Å and α=γ=90$^o$, β=120$^o$. The crystal structures is represented in inset of Figure 1, where Pd is at (0,0,0) with site symmetry -3m and Te at (1/3, 2/3, 1/4) with site symmetry -6mc.

Resistance versus temperature measurements are shown in Figure 2, which clearly indicates metallic normal state with a superconducting onset (T$_c^{onset}$) at 4.5K and T$_c$(R=0) at 4.25K. The T$_c$(R=0) is taken as being function of R, where R becomes zero. The resistance obeys R=R$_0$+AT$^2$ (solid red line in Figure 2) in temperature range of 6K to 50K, suggesting that the dominant scattering at low temperatures is of electron-electron type; a hallmark of Fermi-liquid. The residual resistance ratio (RRR; the ratio of the resistance at room temperature to the one at zero temperature) is estimated to be 21, which is smaller than one being observed for recently studied single crystal of PdTe [12]. As there are no other reports on 4.5K polycrystalline PdTe normal state resistance data, hence we are not been able to compare the RRR value of our sample. Inset of Figure 2 shows the magneto-resistance measured perpendicular to the applied magnetic field up to 1500Oe. With increasing magnetic field, $T_c(R=0)$ decreases at the rate 1.5K/kOe. Based on R=0 criteria, the estimated upper critical field at absolute zero temperature i.e., $H_{c2}(0)=H_{c2}(T)/[1-(T/T_c)^2]$ is 1471Oe, which is well within Pauli paramagnetic limit of $1.84T_c$ [15,16].

The magnetic measurements are carried out to further confirm the superconducting behavior of the polycrystalline PdTe. The dc magnetization is recorded in both zero field cooled (ZFC) and field cooled (FC) conditions for PdTe with 10Oe applied magnetic field, and is shown in Figure 3. A sharp transition at 4.5K with transformation from small positive to negative magnetization for both ZFC and FC cases can be observed, indicating appearance of bulk superconductivity below this temperature. The ac magnetization curves comprising of both the real part M' and the imaginary part M" at 333Hz frequency and varying amplitudes of 3-15Oe amplitude are shown in inset of Figure 3. Both the real (M') and imaginary (M") parts confirm superconductivity in terms of diamagnetic transition in M' and a peak seen in M" [17-19]. It is thus clear from both dc and ac magnetic susceptibility measurements (Figure 3), that PdTe is a bulk superconductor below 4.5K, which substantiates the transport results being shown in Figure 2.

After confirming superconductivity in PdTe at below 4.5K, we study the physical properties in superconducting state i.e., below $T_c$. Figure 4 shows the isothermal magnetization curves recorded from 2K to 4.5K in the superconducting state. As magnetic field increases from zero, the absolute value of magnetization increases linearly up to $H_{c1}$ (=0.25kOe at 2K) suggesting diamagnetic character. Above the magnetic field $H_{c1}$ (marked in Fig.3), the absolute value of magnetization starts decreasing and reaches to zero and become positive above applied field of 1.2kOe at 2K. The similar trend follows at higher temperatures as far as the sample is in superconducting state, i.e., below 4.5K. The magnetization curves clearly confirm the type-II nature of superconductivity with $H_{c1}$ = 250Oe and an upper critical field ($H_{c2}$) of above 1.2kOe at 2K.

The critical fields at absolute zero temperature for a superconductor provide important thermodynamic information. Figure 5 depicts the variation of critical fields; lower critical field ($H_{c1}$) and upper critical field ($H_{c2}$) as a function of normalized temperature i.e., $T/T_c$. Solid lines represent the fitting of lower and upper critical fields to the equation $H_{c1}(T)=H_{c1}(0)[1-(T/T_c)^2]$ and $H_{c2}(T)=H_{c2}(0)[1-(T/T_c)^2]$ respectively. The experimental values of the lower ($H_{c1}$) and upper ($H_{c2}$) critical fields at various temperatures are taken from the isothermal magnetization (Fig. 3) and magneto-resistivity (Fig.2) measurements respectively. We followed the same criteria as used by Karki et al in ref. 12 for the PdTe superconductor. From extrapolation of critical fields to absolute zero temperature one gets $H_{c1}(0)=333 Oe$ and $H_{c2}(0)=1471 Oe$. The small $H_{c2}(0)$ implies a long superconducting coherence length of $\xi_0=449Å$, according to $H_{c2}(0)= \Phi_0/2\pi\xi_0^2$, where $\Phi_0 =2.0678\times10^9 Oe\cdot Å^2$ [19,20]. The thermodynamic critical field $H_c(0)$ can be obtained by arithmetic mean of the upper and lower critical fields at absolute zero temperature i.e. $H_c=(H_{c1}*H_{c2})^{1/2}$ and the resulted value is 700Oe. In Meissner state, using Ginzberg-Landau theory the upper critical field and thermodynamic critical field are related by; $H_{c2}=2^{1/2}\kappa H_c$. Thus estimated Ginzburg–Landau parameter $\kappa$ is $1.48 > 1/2^{1/2}$, implying type-II superconductivity in PdTe [19]. Further, the penetration depth $\lambda(0)$ is calculated from relation $\lambda(0) = \kappa\xi(0)$, which comes out to be 665Å.

The low temperature specific heat probes the low-energy excitations superconducting energy gap i.e., its magnitude and symmetry, which in turn gives the information about ground state of the system. Low temperature molar specific heat $C_p$ of PdTe at different applied magnetic fields is shown in Figure 6. The solid line is the fit to the relation $C_p(T)/T=\gamma+\beta T^2+\delta T^4$ with best fitting for specific heat capacity coefficients; $\gamma=6.01 mJ/mol-K^2$, $\beta=0.93 mJ/mol-K^4$ and $\delta=0.0048 mJ/mol-K^6$, which results in Debye temperature $\Theta_D=(234zR/\beta)^{1/3}$ of about 203K. Here z is number of atoms in PdTe unit cell, which is 4 and R is the Rydberg constant i.e., 8.314J/mol-K. Namely, the $\gamma T$ is the normal state electronic contribution and the $\beta T^3$ and $\delta T^5$ are the lattice contributions to the specific heat. Inset-I of Figure 6 shows the electronic contribution of specific heat $C_e/T$ as a function of temperature. Here $Ce(T) = C_p-\beta T^3-\delta T^5$, i.e. total $C_p$ minus the lattice contribution. The observed specific heat jump $\Delta C/\gamma T_c=1.33$ is lower than the BCS theory value of 1.43, suggesting weak coupling. Interestingly, Karki et al. [12] got a value of 1.67 for their PdTe single crystalline sample thus indicating towards strong coupling. In fact, as rightly mentioned in ref. 12, the specific heat jump ($\Delta C/\gamma T_c$) depends upon superconducting volume fraction and as a result it is difficult to comment on its exact value [19-21].

The electronic heat capacity ($\gamma = C_p/T$), assuming thermal contribution to specific heat does not get affected by magnetic field, can give important information about the nature of superconducting gap with applied magnetic field.. But it is difficult to extract the same accurately from total heat capacity because of dominating phonon contributions. One way to deal with this problem is to measure heat capacity under magnetic field at temperatures much below $T_c$ of a superconductor. Inset-II of Figure 6 presents the heat capacity recorded at 2K, as a function of applied magnetic field, which is clearly not linear, and thus discarding the s-wave superconductivity [20-23]. The solid red line represents the fitting $C_p = a.H^b$, where "a" is proportionality constant and "b" the power factor. The best fitting of heat capacity under magnetic field at 2K for studied PdTe superconductor is obtained at b= 0.22, not following the $H^{0.5}$ behavior as predicted for a developed d-wave superconductor [21,22]. The obtained value of 0.22 though does not follow the exact d-wave pairing, but the non linearity in $C_p(H)$ data at 2K possibly discards the s-wave pairing. Both the $H_{c1}$ and $H_{c2}$ at 2K are marked in $C_p(H)$ plot (inset-II, Figure 6), which are in agreement with the magnetization and transport measurements on PdTe superconductor shown in Figures 2 and Figure 4. Worth mentioning is the fact that the electronic specific heat capacity analysis done in present case is at T = 2K, which is though lower than the $T_c$ = 4.5K of the studied PdTe superconductor, the same in principle is warranted at a temperature of $0.1T_c$ [20-23]. In any case, the $C_p(H)$ data at 2K demonstrate that PdTe is possibly not an s-wave superconductor, and could even be the d-wave one, which needs to be complimented by other experimental techniques.

The density functional calculations on PdTe were carried out using spin-polarized density functional theory within the generalized gradient approximation (GGA) of Perdew, Burke and Ernzehof, using the Vienna Ab-initio Simulation Package (VASP), to calculate the ground state electronic band structure and density of states [24, 25]. The eigen states were expanded in the plane wave basis function and ion cores were considered with the projector augmented wave (PAW) pseudopotentials. The cutoff energy for the plane wave was set at 500eV. The initial structure was taken from Rietveld refined XRD structure and further structure was optimized with force convergences < 0.01eV/Å and total energy convergence of ~ $10^{-6}$eV. The calculated electronic band structure and density of states (DOSs) within GGA approximation are plotted in Fig. 7(a) and 7(b) respectively. The high symmetry points, in reciprocal space, considered for band structure calculations are explained in Fig. 7(a). We observed bands crossing at Γ and K high symmetry points, suggesting the metallic character of PdTe, in agreement with our experimental results. The projected density of states for 4d orbital of Pd and 5p orbital of Te, as shown in Fig. 7(b), contribute almost equally at the Fermi-level and also constitute the majority of total density of states. In addition, Fermi-level of PdTe is close to the local minima of the total density of states $N(E_F)$ as shown in Figure 7(b), which is ~ 1.84 states/eV per unit cell. The contribution of electronic heat has been calculated using $N(E_F)$ and estimated $\gamma_e$, coefficient of electronic specific heat is ~ 2.16 mJ mol$^{-1}$K$^{-2}$. The experimental electronic specific heat coefficient is related to the theoretical one by relation $\gamma=\gamma_e(1+\lambda_{ep})$, where $\lambda_{ep}$ is electron-phonon coupling coefficient and used for calculating electron-phonon coupling coefficient. Thus calculated $\lambda_{ep}$=1.78, for PdTe system, is in general agreement with ref. 12.

In summary, we have synthesized and studied the superconducting properties of polycrystalline 4.5K superconductor PdTe. PdTe crystallizes in layered hexagonal structure with space group P6$_3$/mmc. The metallic normal state conduction at low temperature indicates Fermi-liquid nature. PdTe is a type-II superconductor with Ginzburg–Landau parameter (κ) of 1.48. Worth mentioning is the fact, that this is the first study to our knowledge on superconductivity of polycrystalline PdTe with T$_c$ of above 4.5K. The only other study [12], with T$_c$ of above 4.5K for PdTe is on tiny (15µm) crystallites. Further the C$_p$(H) data at 2K demonstrate that PdTe is possibly not an s-wave superconductor, and could rather be the d-wave one.

**Acknowledgment:**

Authors would like to thank their Director NPL India for his keen interest in the present work. This work is financially supported by *DAE-SRC* outstanding investigator award scheme on search for new superconductors. Reena Goyal thanks UGC, India for research fellowship.

**Figure captions:**

Figure 1 (Color online) Room temperature Reitveld fitted powder XRD pattern of PdTe in which difference (blue line) is minimized between observed (open red circle) and calculated (solid black line) patterns. The position of allowed Bragg reflections are shown as bars (pink). Inset shows the crystal structure of PdTe.

Figure 2 (Color online) Resistance versus temperature plot for PdTe from 300K to 2K, showing a clear superconducting transition at 4.5K. Low temperature resistance is fitted to $R=R_0+AT^2$ (solid red line). Inset presents the magneto-resistance at various applied magnetic fields of up to 1.5kOe.

Figure 3 (Color online) dc magnetization recorded in zero field cooled (ZFC) and field cooled (FC) conditions for PdTe with 10Oe applied magnetic field. Inset presents ac magnetization curves; real M' and imaginary M" at 333Hz frequency and 10Oe amplitude.

Figure 4 (Color online) Isothermal magnetizations (M) in superconducting state (2K to 4.5K) of PdTe as a function of applied magnetic field (H). Inset presents low field one coordinate M-H curve clearly marking the lower critical field ($H_{c1}$).

Figure 5 (Color online) Critical fields; lower critical field, $H_{c1}$ and upper critical field $H_{c2}$ as a function of normalized temperature i.e., $T/T_c$. Solid lines represent the fitting of lower and upper critical fields to the equation $H_{c1}(T)=H_{c1}(0)[1-(T/T_c)^2]$ and $H_{c2}(T)=H_{c2}(0)[1-(T/T_c)^2]$ respectively.

Figure 6 (Color online) Molar specific heat $C_p$ of PdTe recorded at different applied magnetic fields. The solid red line is the fit to the relation $C_p(T)/T=\gamma+\beta T^2+\delta T^4$ Inset-I: change of specific heat $C_e/T$ as a function of temperature, where $C_e(T) = C_p-\beta T^3-\delta T^5$. Inset-II presents the heat capacity recorded at 2K as a function of applied magnetic field and solid red line represents the fitting $C_p=a.H^b$.

Figure 7 (Color online) (a) Calculated electronic band structure along high symmetry points in Brillouin zone for PdTe, where Fermi level is set to 0eV. (b) Total and projected (4d orbital of Pd and 5p orbital of Te atoms) density of states for PdTe when Fermi- level is set to zero.

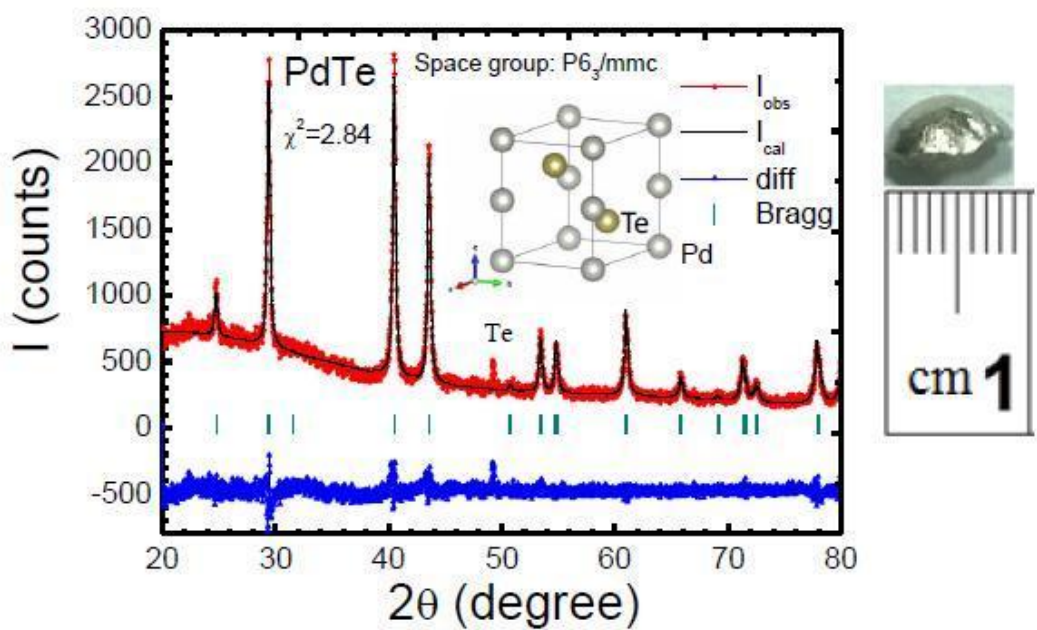

Figure 1

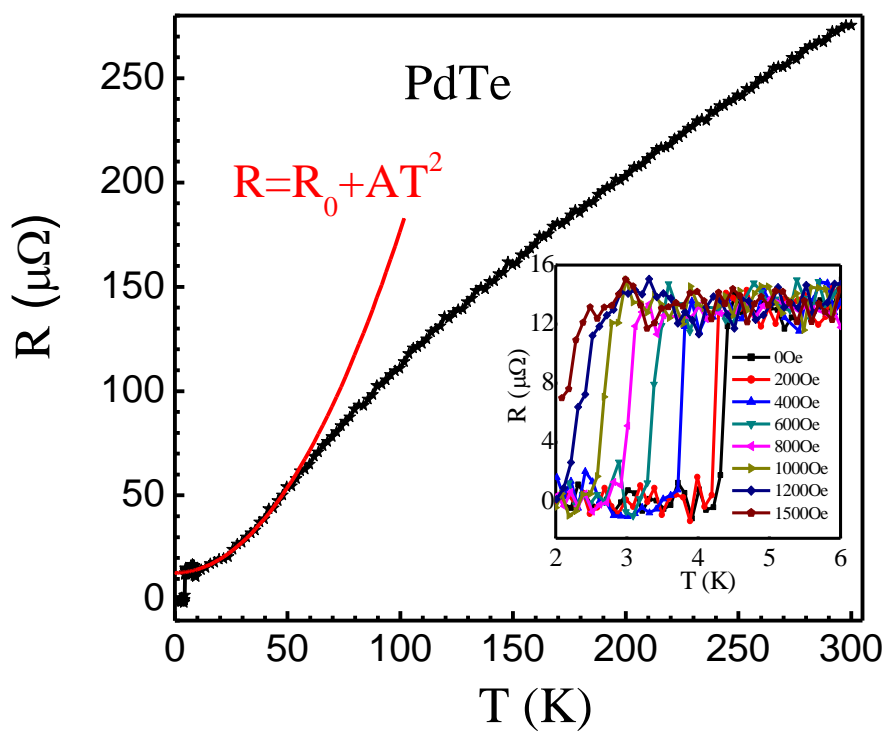

Figure 2

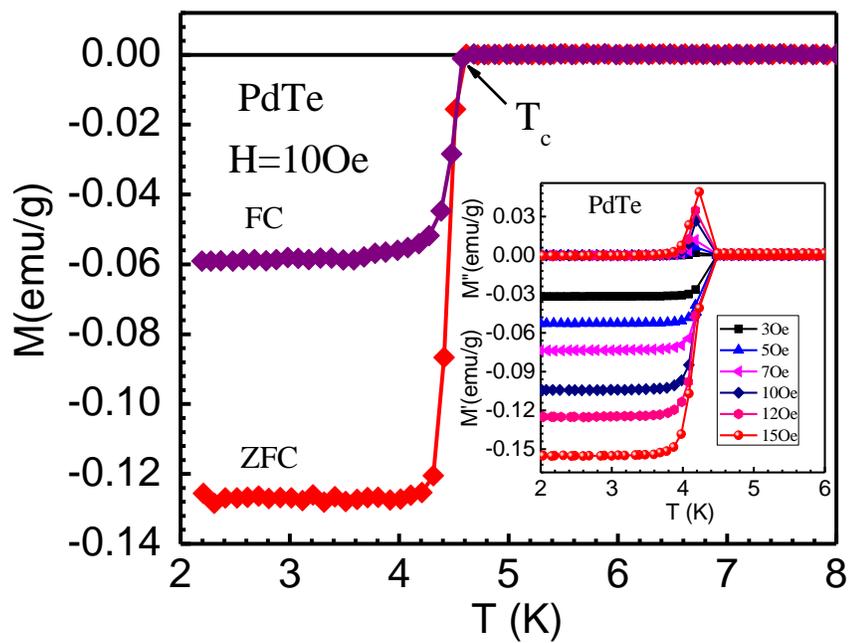

Figure 3

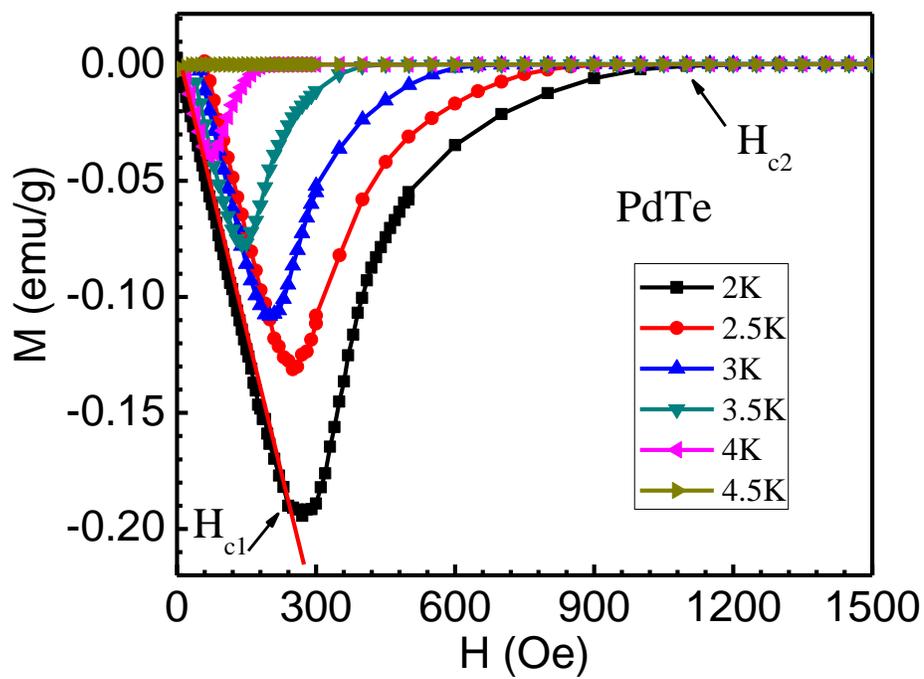

Figure 4

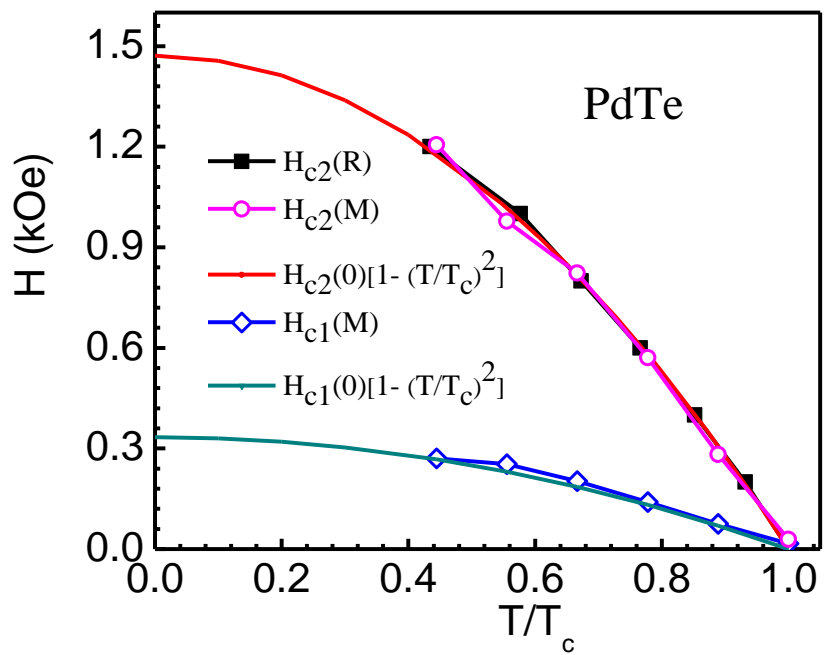

Figure 5

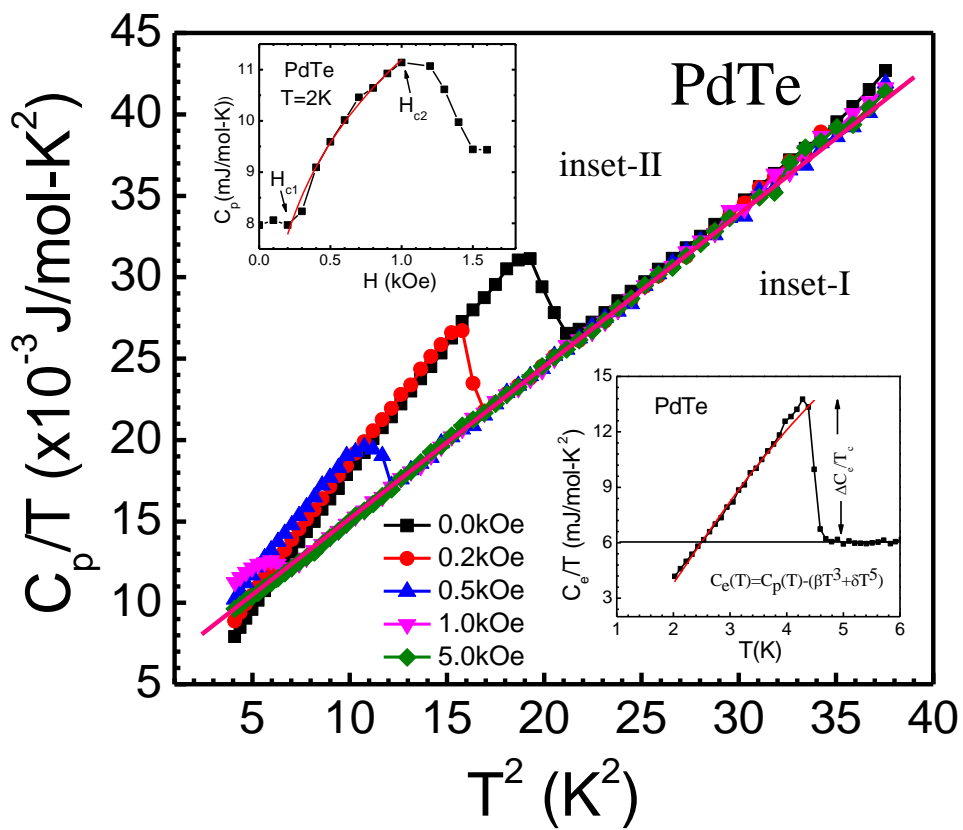

Figure 6

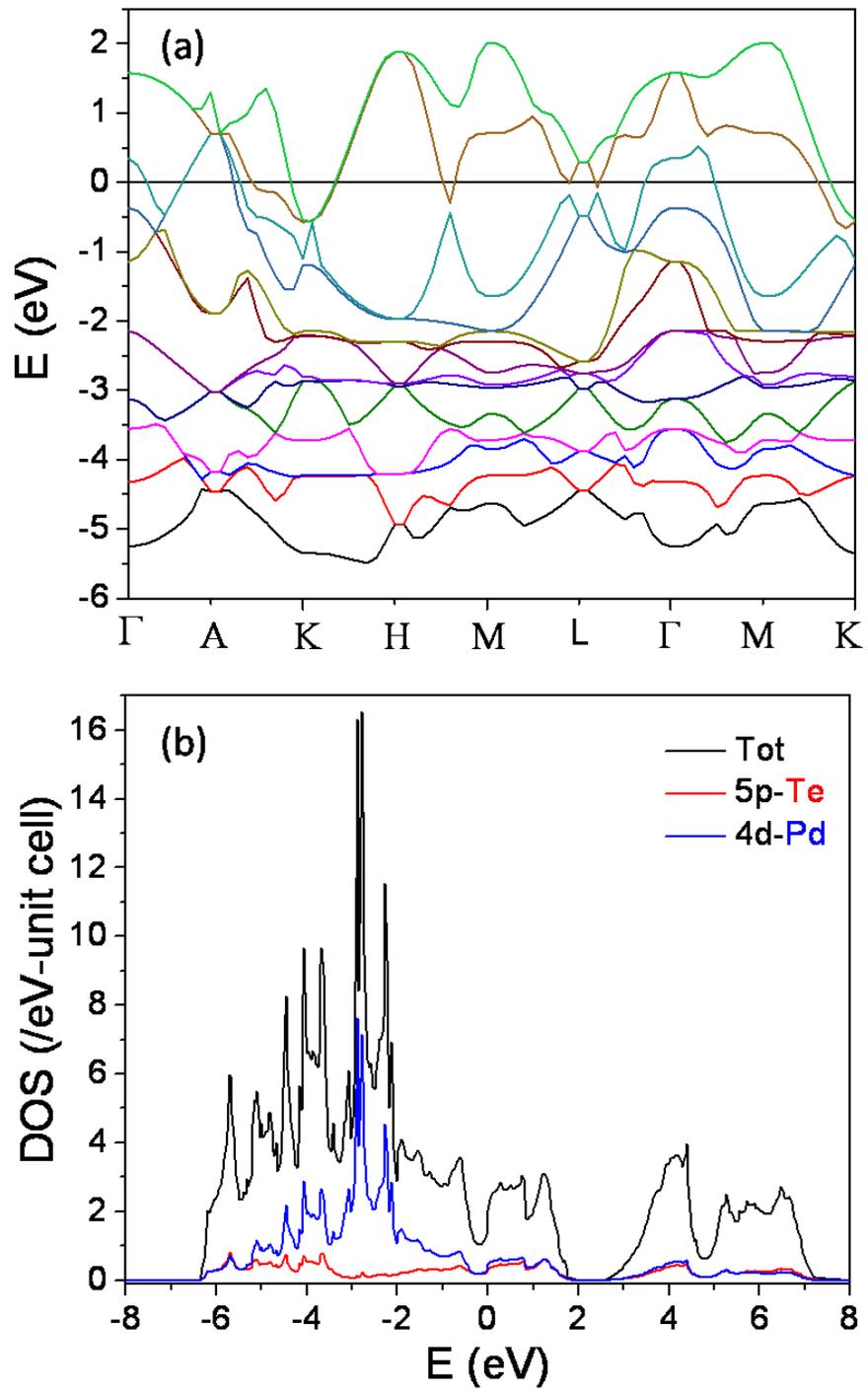

Figure 7